\documentclass[11pt]{article}
\usepackage{erice,epsfig}

\def\be{\begin{equation}}
\def\ee{\end{equation}}
\def\bea{\begin{eqnarray}}
\def\eea{\end{eqnarray}}

\begin{document}
\vspace*{4cm}
\title{ELECTRODYNAMICS AT THE HIGHEST ENERGIES}

\author{SPENCER R. KLEIN}

\address{Nuclear Science Division, Lawrence Berkeley National Laboratory,\\
Berkeley, CA, 94720, USA}

\maketitle\abstracts{At very high energies, the brems\-strahlung and
pair production cross sections exhibit complex behavior due to the
material in which the interactions occur.  The cross sections in dense
media can be dramatically different than for isolated atoms.  This
writeup discusses these in-medium effects, emphasizing how the cross
section has different energy and target density dependencies in
different regimes.  Data from SLAC experiment E-146 will be presented
to confirm the energy and density scaling.  Finally, QCD analogs of
the electrodynamics effects will be discussed.}

\section{Bremsstrahlung, Pair Production and the Formation Zone}

Bremsstrahlung and pair production were described by Bethe and Heitler
in 1934.~\cite{BH} In brems\-strahlung (braking radiation), an electron
with energy $E$ interacts with a target nucleus, and slows down,
emitting a photon with energy $k$ in the process. For interactions
with an isolated atom, for $k\ll E$, the Bethe and Heitler
bremsstralung cross section scales as
\begin{equation}
{d\sigma\over dk} \approx {1\over k}.
\label{eq:bh}
\end{equation}

In pair production, a photon fluctuates to an $e^+e^-$ pair.  The
newly created electron or positron interacts with the electromagnetic
field of a target nucleus, and the pair becomes a real $e^+e^-$ pair.
For photon energies $k \gg m_e$, the pair production cross section is
independent of $k$.  Sophisticated calculations of bremsstrahlung and
pair production confirm that these energy dependencies hold within a
few percent.

Bethe and Heitler treated bremsstrahlung and pair production as
occuring at a single point in space.  With this assumption, the
radiation depends on the change in electron velocity $\Delta\vec{v}$
due to the scattering from the target,
%: $d\sigma/dk \approx |\Delta\vec{v}|^2$, 
independent of the nature of the force that causes the velocity
change. For any interaction, one can determine the expected
$\Delta\vec{v}$ distribution, and, from that, find the bremsstrahlung
radiation.  With this approach, it is relatively easy to generalize
from isolated atoms to a dense medium.

In 1953, Ter-Mikaelian~\cite{priroda} and Landau and
Pomeranchuk~\cite{LP} pointed out that the assumption that the
interaction occurs at a single point fails when the incoming
particle has a high enough energy.  It is impossible to localize the
reaction to a point on the projectile's trajectory.  This is true
classically as well as quantum mechanically.  The pathlength over
which the reaction can be localized is the formation length, $l_f$.

Classically, the formation length is the distance $z$ over which the
phase factor, $\exp(i[\vec{k}\cdot z - \omega t])$ is roughly constant
($\vec{k}\cdot z - \omega t < 1$).  In a vacuum, the photon wave
vector $\vec{k}$ and the photon frequency $\omega$ are related by
$|\vec{k}|c = \omega$. The position $z$ and time are related by the
electron velocity. The electron moves more slowly than the photon, so
eventually the two particles will separate by more than one photon
wavelength, at which point they can be considered two separate
particles.

Quantum mechanically, the formation length is calculated from the
momentum transfer $q_{||}$ from the target nucleus to the
electron-photon system; $l_f =\hbar /q_{||}$.  The momentum transfer
is fixed by the kinematics (4-momentum conservation). For $E\gg m$
with initial ($p_e$) and final ($p_e'$) electron momenta,
\begin{equation}
q_{||} = p_e - p_e' - k = \sqrt{E^2-m^2} - \sqrt{(E-k)^2-m^2} - k
\approx { m^2c^3k \over E(E-k)}
\end{equation}
where $m$ and $E$ are the electron mass and energy.  This holds as
long as the photon emission angle $\theta_\gamma$ is smaller than the
typical emission angle $m/E$.

Because $q_{||}$ drops as $E$ rises, $l_f$ can be very long for high
energy reactions.  For example, for a 25 GeV electron emitting a 100
MeV photon, $l_f = 100\mu$m.  For astrophysical energies, $l_f$ can
be hundreds of meters; for $E=10^{20}$ eV, $k=5\times10^{19}$eV, $l_f
= 160$ m.  As (for $k/E$ fixed) $E\rightarrow\infty$ and
$l_f\rightarrow\infty$, interactions with the medium as a whole
determine the radiation.

For pair production, the formation length is given in a manner similar
to bremsstrahlung.  $l_f = E(E-k)/m^2c^3k = 2\hbar k/(M_p^2c^3)$ where
$M_p$ is the invariant mass of the pair.  This formula also applies to
photoproduction of quark pairs (vector mesons) in hadronic
environments.  

The effects of the formation zone have been discussed in several
reviews.~\cite{TM,review}

\section{Bremsstrahlung Suppression}

Landau and Pomeranchuk~\cite{LP} generalized Bethe and Heitler's
concept of radiation from a scattering at a single point, to radiation
from the total scattering in a single formation zone.  The radiation
cannot be localized within the formation zone, so the entire formation
zone acts as a single emitter, with radiation determined by the
total scattering angle, $\theta_{MS}$ in this length.

\subsection{Suppression due to multiple scattering}

Bremsstrahlung from relativistic particles is reduced when the
scattering angle is larger than $m/E$.  For an isolated interaction,
the usual scattering angle is less than $m/E$, so this is relatively
unimportant. However, in a dense medium, many scatters add (in
quadrature) to a single $\theta_{MS}$, and the mean scattering angle
can be much larger than $m/E$.  When this happens, the radiation is
reduced. The scattering decreases the electron forward velocity
($v_z$), making the classical phase ($\vec{k}\cdot z - \omega t$)
vary more rapidly with $z$.

\begin{figure}
\center{\psfig{figure=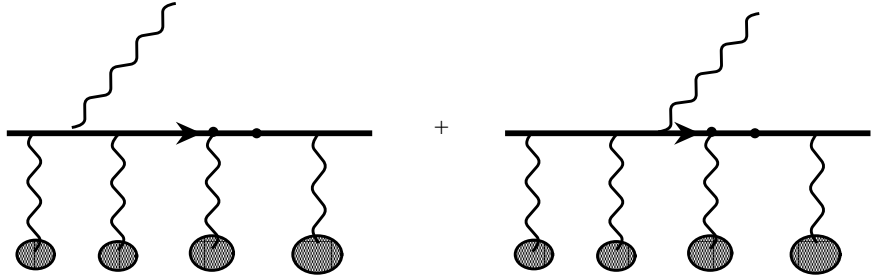,height=1.6in,clip=}}
\caption{Two Feynman diagrams for bremsstrahlung in a dense medium.
If the nuclei are packed densely enough, the amplitudes for these two
processes nearly cancel.}
\label{fig:diagram}
\end{figure}

This reduction can also be explained quantum mechanically.
Fig. \ref{fig:diagram} shows the radiation from two pieces of the
electron trajectory, separated by a target nucleus.  When the nuclei
are close enough together (separation much less than $l_f$), the
amplitudes for emission from adjacent electron lines are almost
completely out of phase, and so largely cancel. The target can be
thought of as being divided into a small number of independent
radiators, each $l_f$ thick.  In a 1956 work, Migdal treated the
multiple scattering using the Fokker-Planck equation, and calculated
the radiation for each possible path.~\cite{migdal} This approach gave
accurate results for the entire $k/E$ range.  However, for pedagogical
purposes, this contribution will follow the Landau and Pomeranchuk
approach.

Bremsstrahlung is suppressed when $\theta_{MS} = (E/E_s)^2 l_f/X_0$ is
larger than $m/E$.  Here $E_s=mc^2\sqrt{4\pi/\alpha} = 21.2$ MeV,
$X_0$ is the material radiation length, and $\alpha\approx 1/137$ is
the fine structure constant.  Defining
\begin{equation}
E_{LPM} = {m^4c^7 X_0 \over \hbar E_S^2} = 7.7\ {\rm TeV/cm}\cdot X_0,
\end{equation}
we find $\theta_{MS} >m/E$ when $k/E < (E-k)/E_{LPM}$.  The critical
energy, $E_{LPM}$ ranges from few TeV in solids (4.3 TeV in lead, 2.5
TeV in uranium) up to 540 TeV in water, and 234 PeV in air at sea
level.  For bremsstrahlung, of course, suppression is present for all
$E$, as long as $k/E < E/E_{LPM}$.  This reduction holds for photons
emitted at small angles ($\theta < m/E$).  Large angle photon emission
( $\theta\gg m/E$) already requires a large $q_{||}$ and is much less
suppressed.

\begin{figure}
\center{\psfig{figure=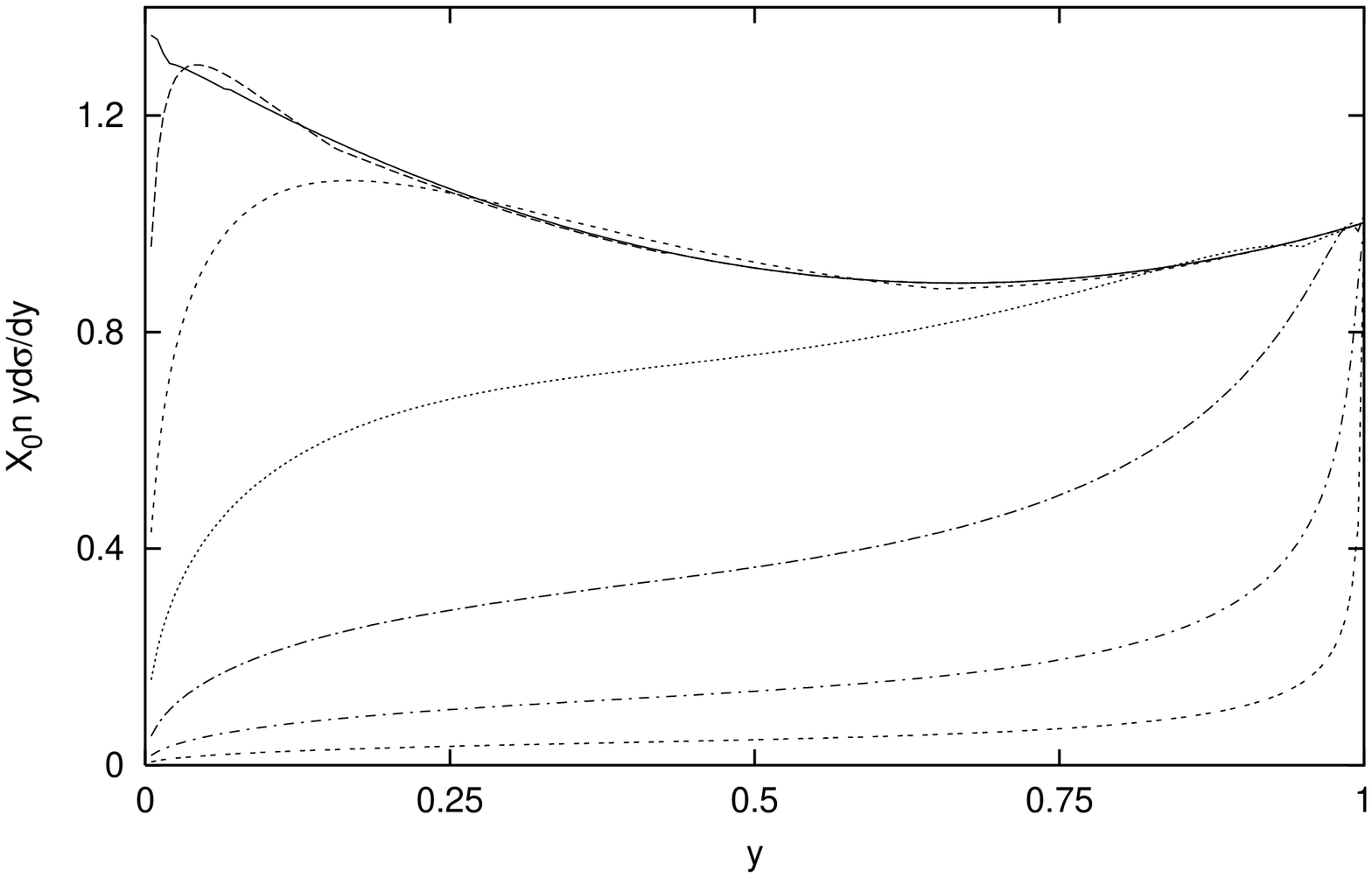,height=3.05in,angle=270,clip=}
{\psfig{figure=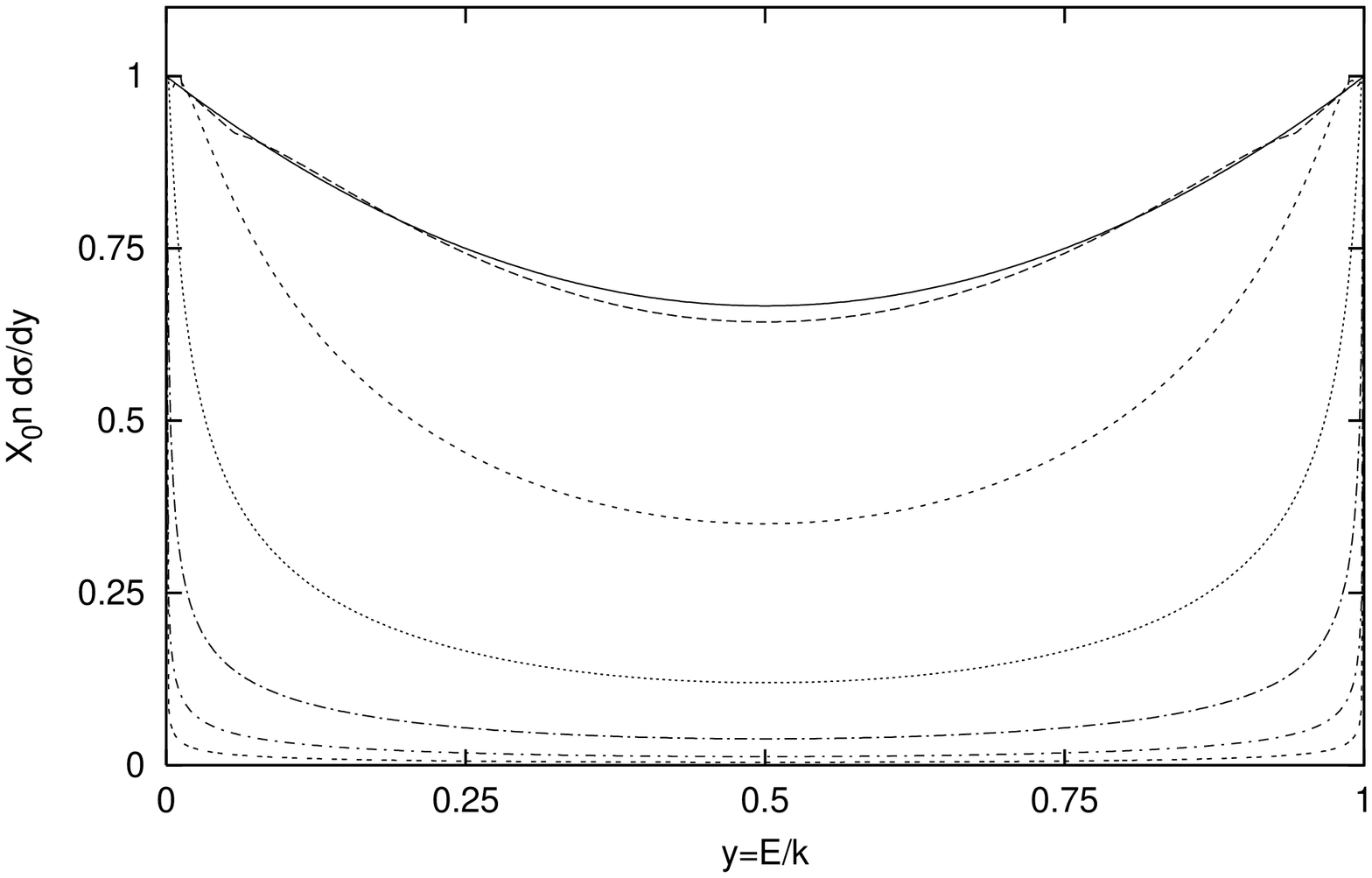,height=3.05in,angle=270,clip=}}}
\caption{(a) The energy weighted differential cross sections
$yd\sigma/dy$ ($y=k/E$) for bremsstrahlung for electrons with energies
$E=10$ GeV, 100 GeV, 1 TeV, 10 TeV, 100 TeV, 1 PeV and 10 PeV in lead;
as $E$ increases, the cross section decreases. The units are
fractional energy loss per radiation length. (b) The differential
cross sections $yd\sigma/dy$ (Here $y=E/k$) for pair production for
photons with energies $E=1$ TeV, 10 TeV, 100 TeV, 1 PeV, 10 PeV 100
PeV and 1 EeV in lead; as the energy increases, the cross section
decreases. The units are fractional energy loss per radiation length.}
\label{fig:bremscaling}
\end{figure}

In the regime of strong suppression, the bremsstrahlung cross section
is
\begin{equation}
{d\sigma\over dk} \approx {1\over\sqrt{k}};
\label{eq:lpm}
\end{equation}
the photon energy dependence changes.  The change is due to the
reduced coherence. The amplitudes for each bit of path-length $dz$ in
$l_f$ add in-phase, so the cross section $\sigma = |\int A(z) dz|^2$.
If a $z-$dependent phase reduces the coherence, the cross section is
also reduced, even if $|A|$ is unchanged.  A calculation of $l_f$
including the additional momentum transfer required due to the
multiple scattering will yield Eq. (\ref{eq:lpm}).
Fig. \ref{fig:bremscaling}(a) shows the cross section following
Migdal's calculations.  In the limit $E\gg E_{LPM}$, suppression is
near total except for $k\approx E$.

Pair production is suppressed if the produced electron and positron
multiple scatter enough. This may be calculated from the bremsstahlung
formulae using the crossing symmetry that relates the two processes.
Fig. \ref{fig:bremscaling}(b) shows the pair production cross
sections.  Small $M_p$, symmetric pairs are suppressed the most; large
$M_p$ asymmetric pairs have a naturally small $l_f$ and are less
subject to suppression.  Suppression is significant for $M_p^2 <
km^2/E_{LPM}$.  Of course, since $M_p \ge 2 m$, suppression requires
$k> 4E_{LPM}$.  When suppression is strong, the pair production cross
section scales as
\begin{equation}
\sigma(k) \approx {1\over \sqrt{k}};
\end{equation}
the cross section decreases as $\sqrt{k}$, similar to the
bremsstrahlung case. For $k\gg E_{LPM}$, very asymmetric pairs are
created.

\subsection{Suppression due to photon Compton scattering}

Photons from bremsstrahlung are also affected by their environment.
The photons can Compton scatter off the electrons in the medium.  The
collective forward Compton scattering introduces a phase shift in the
photons.  The shift can be described classically using the dielectric
constant of the medium:
\begin{equation}
\epsilon(k) = 1 - \big( {\hbar\omega_p\over k} )^2
\end{equation}
where $\omega_p$ is the plasma frequency of the medium;
$\hbar\omega_p$ is typically 40-60 eV for solids.  Now, $|\vec{k}|c
=\sqrt{\epsilon} \omega$, giving the photon an effective mass
$\hbar\omega_p$.  This effective mass ($\epsilon \ne 1$) introduces a
phase shift, reducing $l_f$.  The reduction is significant when $k <
\gamma\hbar\omega_p$.  This inequality is satisfied when $k/E <
\hbar\omega_p/mc^2\approx 10^{-4}$. The latter number applies for
solids.  For smaller $k$, the photon mass dominates, and $l_f =2ck
/\hbar\omega_p^2$ and the cross section scales as~\cite{TM}
\begin{equation}
\sigma \approx k.
\end{equation}
The $k-$dependence is drastically changed, and the radiation
disappears as $k\rightarrow 0$.  This is sometimes known as the
longitudinal density effect or as dielectric suppression.  Migdal
allowed for dielectric suppression in his calculations, allowing the
combined effect to be calculated

As $k\rightarrow 0$, $q_{||}$ increases and $l_f$ decreases, actually
increasing the localization.  Baier and Katkov pointed out that, for
sufficiently small $k/E$, the interaction may be localized to within
the atomic nucleus, and the nuclear form factor affects the
emission.~\cite{bkat}

Because of this suppression, the infrared divergence completely
disappears!  The medium eliminates the need for artificial
cutoffs. The total bremsstrahlung cross section is then finite.  For
example, a 25 GeV electron in lead emits about 10 photons per
radiation length.

\subsection{Mutual Suppression of Bremsstrahlung and Pair Production}

\begin{figure}
\center{\psfig{file=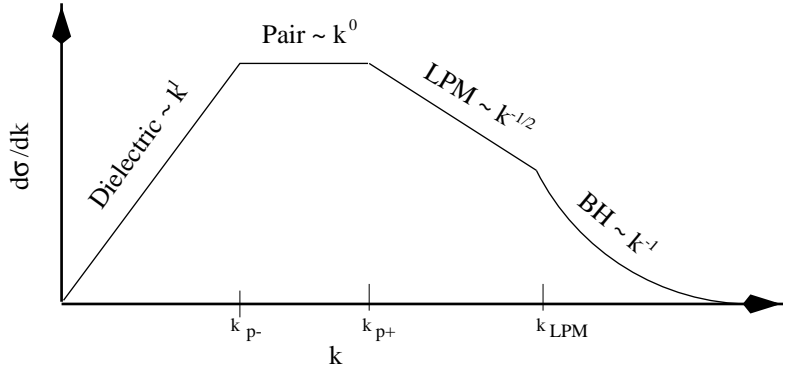,height=2.1in,%
bbllx=0,bblly=0,bburx=520,bbury=270,%
clip=,angle=0}}
\vskip .1 in
\caption[]{Schematic view of bremsstrahlung $d\sigma/dk$ with several
suppression mechanisms, showing the different $k$ ranges.  For $E<
E_p$, the pair creation suppression disappears and LPM suppression
connects with dielectric suppression.}
\label{bremspectrum}
\end{figure}

At sufficiently high energies, $l_f > X_0$.  Then, a nascent photon
from a bremsstrahlung interaction will convert to an $e^+e^-$ pair
before it is fully formed.  The pair production is localized (to
within it's own $l_f$), and can be used to better locate the
bremsstrahlung.  This localization reduces the formation length.  In
the simplest approach one can limit the formation length to
$X_0$.~\cite{galgu} Then $\sigma/\sigma_{BH} = X_0/l_f$ and
\begin{equation}
{d\sigma\over dk} = k^0
\end{equation}
{\it i.e.} $d\sigma/dk$ is independent of $k$.  After accounting for
LPM and dielectric suppression, this effect appears for $E> E_p =
X_0\omega_p E_s/\sqrt{2}c$, with $E_p$ about 24 TeV in lead, rising to
205 TeV in aluminum, 540 TeV in water and 15 PeV in sea level air. It
appears for intermediate energy photons, as is shown in
Fig. \ref{bremspectrum}.

A similar effect occurs for pair production. One of the produced
leptons will emit a brems\-strahlung photon, thereby localizing the
pair production and reducing $l_f$.

A more sophisticated calculation would treat the bremsstrahlung and
pair production as a single Feynman diagram, finding the amplitude for
the combined interaction.  As the incident energy rises, the
amplitudes for more and more generations of the electromagnetic shower
become intertwined, and the calculation become intractable.

\subsection{Finite Targets and Surface Radiation}

The calculations discussed so far only apply to infinite targets.
Finite thickness targets have surfaces which can affect the radiation.
The simplest case is a very thin target, with thickness $T$ less than
$l_f$.  Then, the target interacts as a whole, and the radiation
depends on the total scattering in the target:~\cite{shulga}
\begin{equation}
{dN\over dk} \approx \ln{\big({1720T\over X_0}\big)} -1.
\label{eq:total}
\end{equation}
Here, $dN/dk$ is the radiation from the entire target.  The radiation
depends only logarithmically on the target thickness!  It is important
to accurately model for the Coulomb scattering.  The usual Gaussian
approximation underestimates the number of large angle scatters, and
hence the radiaton.  Eq. (\ref{eq:total}) is based on an an accurate
distribution. The radiation will vary from electron to electron,
depending on the scattering.  Electrons that scatter a lot will emit
more radiation than those that scatter less.

For thicker targets, it is conceptually simple to consider the
radiation as coming from a bulk target plus the entrance and exit
surfaces.  This surface radiation is naturally treated as a type of
transition radiation.  As long as absolute rigor is not required, this
is a fruitful approach.

Conventional transition radiation is due to the adjustment that an
electrons electromagnetic fields make when they enter a medium with a
different dielectric constant.  The fields must make a comparable
adjustment when the electron enters a medium with a different amount
of multiple scattering per unit length; the multiple scattering causes
the electron to jitter and the electron drags the fields with it.
Ter-Mikaelian pointed out that both types of transition radiation
depend on the difference in formation length between the two
media:~\cite{TM}
\begin{equation}
{dN \over dk} \approx (l_f -l_f')^2.
\end{equation}

\begin{figure}
\center{\epsfig{file=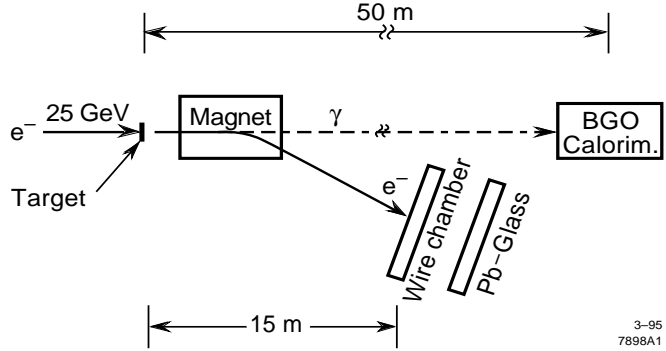,height=2.in,width=3.65in,%
bbllx=145,bblly=310,bburx=420,bbury=470,%
clip=,angle=0}}
\caption{A block diagram of SLAC Experiment E-146 in End Station A.
Electrons traversed a thin target and were bend downward into a
set of wire chambers and lead glass blocks.  Bremsstrahlung photons
travelled 50 m downstream into a BGO calorimeter.}
\label{fig:experiment}
\end{figure}

In the 1960's, several calculations of this transition
radiation appeared. Although the basic methods were similar,
the numerical results were not.

Newer calculations have treated the target as an integrated whole,
with good results.  These calculations incorporated many advances; a
fuller description and references is given in Ref. 5.  Almost all of
the calculations allowed for electrons traversing arbitrary density
profiles.  The path is considered as a whole, including interference
between radiation from each pair of points on the trajectory. Most of
the calculations used accurate models of the Coulomb potential instead
of the Gaussian scattering model.  Both elastic and inelastic
(electron-electron) scattering were included.

\subsection{Experimental Tests}

Shortly after Migdals work appeared, several groups tried to measure
LPM suppression of pair conversion from very high-energy photons in
cosmic rays.  These early experiments all suffered from very limited
statistics, typically 1-50 events.  The first accelerator experiment,
at Protvino in 1976, used 40 GeV electrons interacting in solid
targets.~\cite{protvino} Their data indicated that there was some
suppression.  Later, Kasahara studied the development of $\approx$100
TeV showers in lead/emulsion chambers, and found a significant
elongation, matching the LPM predictions.~\cite{kasahara} This study
is of special interest because it is the only experiment to probe the
region $E\approx E_{LPM}$.

The first accurate measurement was by SLAC experiment
E-146,~\cite{e146lpm,e146diel} in 1993.  The collaboration studied
bremsstrahlung of 200 keV to 500 MeV photons from 8 and 25 GeV
electrons in 7 different target materials.  At least two different
thicknesses were used for each target material.  The target
thicknesses were optimized to trade off between multiple interactions
in thick targets vs. transition radiation in thin targets. As
Fig. \ref{fig:experiment} shows, electrons interacted in a thin target
and were magnetically bent downward into a set of momentum-measuring
wire chambers and electron-counting lead glass blocks.  A 45 crystal
array (7 by 7, minus the 4 corners), 18 $X_0$ thick BGO calorimeter
detected the emitted photons The calorimeter segmentation was used to
determine the photon position.

The collaboration took great care to minimize systematic errors.  The
experiment took data at 120 Hz, and only bunches containing a single
electron were used for the analysis. The electron beamline upstream of
the target was kept in vacuum to minimize bremsstrahlung from air.
The calorimeter was calibrated by two independent methods. 

Semi-independent analyses were performed for $k>5$ MeV and $k<5$ MeV
photons.  For $k>5$ MeV, most photon interactions in the calorimeter
were pair conversions.  At lower energies, Compton scattering
dominated.  Also, for the 25 GeV beams, synchrotron radiation from the
bending magnets was significant for $k<$ 1 MeV.  Most of this
radiation was from the downward bend by the spectrometer magnet, with
a much smaller contribution from the beam-line steering magnets (not
shown).  The contribution from the spectrometer magnet was largely
removed by dividing the calorimeter into diagonal quadrants.  Data
from the bottom quadrant, where most of the synchrotron radiation hit,
was not used.  For $k>5$ MeV, the systematic error was 5\%.  Below 5
MeV, the systematic error was 9\% for the 8 GeV electrons, rising to
15\% for the 25 GeV electrons.  The increase is because of the
synchrotron radiation removal cut, because the cut efficiency depends
on how well the beam is centered.

\begin{figure}
\center{\psfig{file=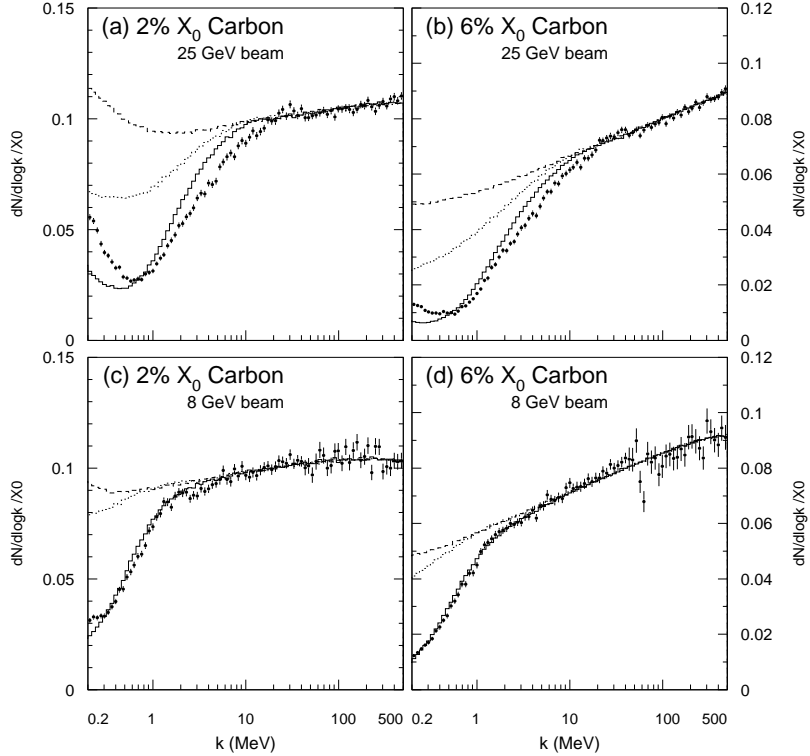,width=4.25 in,%
bbllx=30,bblly=165,bburx=570,bbury=670,%
clip=,angle=0}}
\vskip .1 in
\caption[]{SLAC-E-146 data for 200 keV to 500 MeV photons from 8 and
25 GeV electrons passing through carbon targets. The cross sections
are given as $dN/d(\ln k)/X_0$ where $N$ is the number of events per
photon energy bin per incident electron, for (a) 2\% $X_0$ carbon and
(b) 6\% $X_0$ carbon targets in 25 GeV electron beams, while (c) shows
the 2\% $X_0$ carbon and (d) the 6\% $X_0$ carbon target in an 8 GeV
beam.  Three simulations are shown.  The solid histogram shows LPM and
dielectric suppression of bremsstrahlung, plus conventional transition
radiation.  Also shown are the Bethe-Heitler plus transition radiation
(dashed histogram) and LPM suppression only plus transition radiation
(dotted histogram) simulations.}
\label{carbon}
\end{figure}

\begin{figure}
\center{\epsfig{file=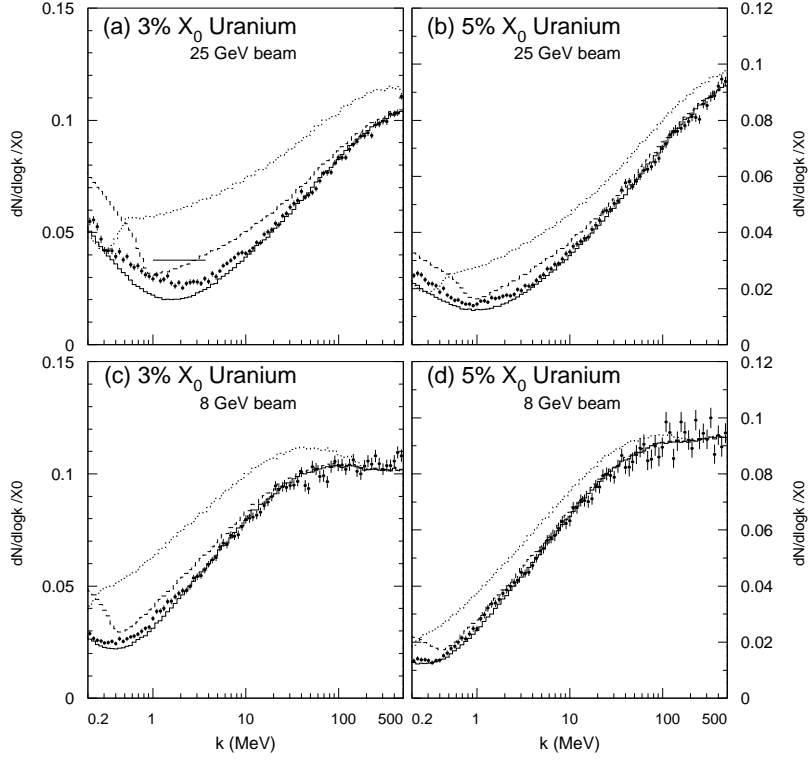,width=4.25in,%
bbllx=30,bblly=165,bburx=570,bbury=670,%
clip=,angle=0}}
\caption[]{SLAC E-146 measurements and simulations for 3\% $X_0$ and
5\% $X_0$ uranium targets in 8 and 25 GeV electron beams.  The solid
histogram shows a simulation with LPM and dielectric suppression and
conventional transition radiation. The other simulations use different
calculations of transition radiation due to multiple
scattering.~\cite{e146lpm} The solid line in panel (a) is based on
Eq. (\ref{eq:total}).}
\label{uranium}
\end{figure}

\begin{figure}
\center{\psfig{file=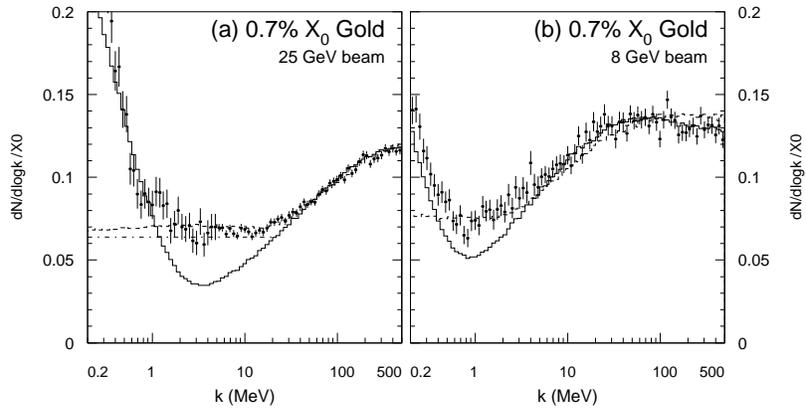,width=4.25in,%
bbllx=30,bblly=160,bburx=572,bbury=446,%
clip=,angle=0}}
\caption[]{SLAC E-146 data on 8 and 25 GeV electrons hitting a
0.7\%~$X_0$ gold target.  Shown are calculations by Blankenbecler and
Drell (dashed line), and using Eq. (\ref{eq:total}) (dot-dashed line).
The solid histogram is a simulation assuming Migdal plus dielectric
suppression.}
\label{gold}
\end{figure}

Some of the E-146 cross section ($d\sigma/dk$) data is shown in
Figs. \ref{carbon} -\ref{gold}.  The histogram bin widths are
proportional to $\ln{(k)}$. With logarithmic bins, the Bethe-Heitler
$d\sigma/dk\approx 1/k$ becomes $d\sigma/d\ln{(k)}$ which does not
vary with $k$. The Bethe-Heitler simulations (dashed lines) in
Fig. \ref{carbon} are sloped because a single electron traversing a
target may interact more than once; the multiple interaction
probability increases with the square of the target thickness. Figure
\ref{carbon} shows 3 simulations: one based on the Bethe-Heitler cross
section, one with LPM suppression alone, and the last with LPM and
dielectric suppression.  Both mechanisms are required to explain the
data.  For the 25 GeV electrons, the upturn in the simulations for $k<
500$ keV is due to conventional transition radiation.  The data has a
similar, but larger upturn.  This may be due to some remaining
synchrotron radiation from the electron bending magnets.

For the 25 GeV carbon data, the simulation and the data are not in
complete agreement for $k<15$ MeV.  This discrepancy is puzzling.
The amorphous graphite target had a micro-crystalline
structure, with non-uniform density at a length scale comparable to
$l_f$.  However, if this were a problem, it should also have
appeared in the 8 GeV data.

For heavier targets, the suppression is much larger and the agreement
is better.  For example, Fig. \ref{uranium} compares the E-146 uranium
target data with an LPM plus dielectric suppression simulation (solid
line). For $k >5$ MeV, the agreement is excellent for both targets at
both 8 and 25 GeV.  At lower energies, the data is above the
simulations.  The difference could be due to the additional transition
radiation discussed earlier.  The other curves in the figure show
different predictions of transition radiation; none fit the data.  The
other targets are believed to be homogenous, with the possible
exception of thin oxide layers on the surfaces.

The collaboration also studied thin targets. Figure \ref{gold} shows
the data for a 0.7\% $X_0$ ($3.2\mu$m) thick gold target.  For 25 GeV
electrons in this target, $l_f > T$ for 1 MeV $< k < 7$ MeV. In this
range the data is considerably above the LPM plus dielectric
prediction.  However, calculations using the thin-target limit,
Eq. (\ref{eq:total}), or based on whole-target approaches work well;
the dot-dashed line in the figure is based on Eq. (\ref{eq:total})
while the dashed line is a whole-target calculation by Blankenbecler
and Drell.

In short, for $E\ll E_{LPM}$, the LPM effect seems to be understood to
the 10\% level in a variety of regimes.  For higher energies, the
effect is clearly there, but good measurements are needed to confirm
the formulae.

\section{QCD Analogs}

The problem of hadrons traversing a hadronic medium is closely related
to that of electrons traversing a charged medium.  Reactions involving
small momentum transfer are poorly localized, and the overall
environment must be considered.~\cite{qcd} This is true at both the
hadronic and partonic level.  Here, we will consider two very diverse
examples: a quark or gluon traversing a nuclear environment, and
vector meson photoproduction in a nucleus.

Vector meson photoproduction is a fairly well studied process, with
many similarities with $e^+e^-$ pair conversion.~\cite{oldrev} A
photon fluctuates to a virtual $q\overline q$ pair.  The $q$ or
$\overline q$ interacts elastically with the target nucleus, and the
pair emerges as a real vector meson.  The elastic scattering can be
modelled by Pomeron exchange.  The details are unimportant here; our
interest is in the target atomic number ($A$) dependence. The nuclear
radius $R_A\approx 1.2A^{1/3}$ fm and $l_f = 2\hbar k/M_V^2$, where
$M_V$ is the vector meson mass.  For $\rho$ production in lead, $l_f >
R_A$ for $k> 20$ GeV, and the nucleus interacts as a whole.  In this
regime, one can further differentiate two regimes.  For vector mesons
where the $q\overline q$ scattering cross section is small ($J/\psi$
or $\Upsilon$), the pair will likely undergo a single interaction in
the nucleus.  Then, the interaction amplitudes from the different
nucleons add coherently, and $\sigma\approx A^2$.  If the $q\overline
q$ dipole is large (the $\rho$), then a single pair can interact with
many nucleons.  As with the LPM effect, the interference is
destructive, and the effects of the interior nucleons disappear.  In
effect, the $q\overline q$ pair sees only the front surface of the
nucleus, with area $A^{2/3}$, and $\sigma\approx A^{4/3}$.  At lower
$k$, where $l_f\ll R_A$, the interactions are independent, and
$\sigma\approx A$.  So, simple kinematics considerations lead to 3
different $A-$scaling regimes, independent of the details of the
interaction.

In contrast, quark or gluon (parton) energy loss in dense media is
still not well understood. A parton produced in a relativistic heavy
ion collision will interact and lose energy to the surrounding medium,
be it hot nuclear matter or a quark gluon plasma.~\cite{anrev} For
most of the energy-loss interactions, $l_f>R_A$ and interference between
multiple interactions is important.  The previously discussed 'thin
target' approach applies: the energy loss should be calculate based on
the total scattering angle.

The created partons are bare, without an accompanying colored fields.
Because of the retarded potentials, newly created partons (or
electrons) will take some time to grow accompanying colored (or
electromagnetic) fields. Until these fields are fully formed, the
interaction cross sections will be reduced.  Because of these effects,
the total energy loss grows as $L^2$, where $L$ is the distance
traversed in the medium.

One key question concerns where the lost energy goes.  Experiments
reconstruct jets by measuring energy deposited in a cone.  If the
energy lost by the parton is transfered to soft particles which remain
in the jet cone, the measured jet energy will be unaffected.  The
energy loss can only be measured by studying the energy of the
individual particles in the jet.  However, if the soft particles
scatter outside the jet cone, the measured jet energy will be
unchanged.  Bremsstrahlung suppression is angular; photons emitted
with angles $\theta_\gamma \gg m/E$ are much less suppressed.  So, in
the strong suppression limit, $\theta_\gamma \gg m/E$.  Because the
angular dependence is entirely due to kinematics, the same effect
should apply in QCD.  When suppression is strong, radiated gluons are
more likely to be outside the jet cone.

Unfortunately, although there is broad agreement on many elements of
energy loss calculations, different numerical results may vary
widely. However, despite the uncertainty, simple kinematic arguments
give significant information about the dependence on $L$ and density.

\section{Conclusions}

High-energy electrodynamic reactions are highly sensitive to their
environment.  In-medium effects can drastically change the cross
sections of bremsstrahlung and pair production.  Individual target
atoms are indistinguishable, and the electron or photon projectiles
interact with the medium as a whole.  At the highest energies, it
becomes impossible to separate electromagnetic showers into individual
reactions, and current theoretical approaches break down.

Similar effects occur for hadronic reactions in dense media such as
heavy ions.  In some cases, kinematics-derived scaling laws can
explain the in-medium effects, while in others, 
more work is needed.

\section*{Acknowledgments}
The author would like to thank Bill Marciano and Sebastian White for
organizing an enjoyable conference. This work was supported by the
U.S. D.O.E. under contract DE-AC03-76SF00098.

%\section*{Appendix}

\end{document}